Number: T093375/FI58544/1

# DIPLOMA THESIS

BOLARINWA OLAYEMI SAHEED

2021

University of Pécs
Faculty of Engineering and Information Technology
Computer Science Engineering Master

# DIPLOMA THESIS

Kernelization of Discrete Optimization Problems on

Parallel Architectures

Author: BOLARINWA OLAYEMI SAHEED

Supervisor: Professor Dr. Sándor Szabó

Pécs

**2021**

# ABSTRACT

There are existing standard solvers for tackling discrete optimization problems. However, in practice, it is uncommon to apply them directly to the large input space typical of this class of problems. Rather, the input is pre-processed to look for simplifications and to extract the core subset of the problem space, which is called the Kernel. This pre-processing procedure is known in context of parameterized complexity theory as Kernelization.

In this thesis, I implement parallel versions of some Kernelization algorithms and evaluate their performance. The performance of Kernelization algorithms is measured either by the size of the output Kernel or by the time it takes to compute the kernel. Sometimes the Kernel is the same as the original input, so it is desirable to know this, as soon as possible. The problem scope is limited to a particular type of discrete optimisation problem which is a version of the K-clique problem in which nodes of the given graph are pre-coloured legally using k colours.

The final evaluation shows that my parallel implementations achieve over 50% improvement in efficiency for at least one of these algorithms. This is attained not just in terms of speed, but it is also able to produce a smaller kernel.

| UNIVERSITY OF PÉCS | Number: T093375/FI58544/1 |
| --- | --- |
| FACULTY OF ENGINEERING AND INFORMATION TECHNOLOGY | 07/06/2021 |

**Computer Science Engineering Master**

# DIPLOMA THESIS

**Bolarinwa Olayemi Saheed**

..................................................................................

**for student**

The title and the topic of the diploma thesis, which must be submitted before the final exam, are the following:

**Title:** Kernelization of Discrete Optimization Problems on Parallel Architectures

**Tasks:** Implementation of parallel kernelization algorithms and evaluate their performance.

Responsible department: Department of System and Software Technology

Supervisor:   Prof. Dr. Sándor Szabó
Institution:   University of Pecs

Pécs, 2021.06.07

<div style="text-align:right">

Prof. Dr. Péter Iványi

Head of department

</div>

# Declaration

I declare that this diploma thesis is the result of my own work. All references and external works have been identified and cited. I have not used any other external help.

The results of my diploma thesis can be used by the university for its own purposes free of charge.

Pécs, 2021.06.07

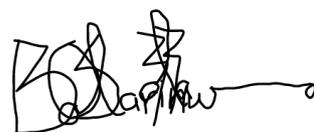

..................................................
signature of the student

# Table of Contents

# 1  Introduction

## 1.1  Overview

In discrete optimisation problems we aim to find the optimal solution among a large solution space which expressed as an objective function with a couple of criteria. These problems are known be ridiculously hard to solve. The k-Clique problem falls within this category and even horned within the category has been listed among Karp's 21 NP-complete problems.[1] The k-Cliques problems show up in several problems that can be modelled using graphs. It has applications in social networks analysis [2], electronic circuit designs [3], bioinformatics [4], as well as many computational science problems. There are existing standard solvers for tackling discrete optimization problems. However, in practice, it is uncommon to apply them directly to the large problem space associated with these kinds of problems. Rather, in practise, the input space is pre-processed in a search for simplifications and the core subset of the problem space is extracted. This core is called the Kernel. This pre-processing procedure is known in parameterized complexity theory as Kernelization.

This thesis examines some Kernelization algorithms to see if we can improve their performance by harnessing the parallelism in today's processing units. The performance of Kernelization algorithms is measured either by the size of the output Kernel or by the time it takes to compute the kernel. When the output size is small, we save compute for the actual solver. Sometimes the Kernel is the same as the original input or almost, so it is desirable to know this, as soon as possible. The holy grail will be to obtain a relatively small kernel within a small period time. However, most kernelization algorithm are either efficient in time or size.

## 1.1  Motivation

There is a growing interest in kernelization as pre-processing technique. The thesis hopes to further extend the feasibility of kernelization algorithms by exploring the use of parallelism. I hope to be able to show that kernelization algorithms can take advantage of massively parallel compute units to improve not just the speed but also the size of the kernel.

## 2.  Literature Review

## 2.1 Discrete Optimisation

Discrete Optimisation problems belong to the class of Optimisation in which the variables are discrete. Mathematical Optimisation (or Mathematical Programming) aims to find the best solution to a defined problem from a given set of feasible solutions to the problem. A set of criteria modelled as objective functions guides the search for optimal solution among the feasible solution space. The literature often interchanges the term "Optimisation" and "Programming" in this context. The use of the term "Programming" thus is not synonymous to act of writing computer programs, though their correlations cannot be ignored.

There are several perspectives from which optimisation problems can be classified [5]. However, optimisation problems are mostly classified by the nature of the variables used in their objective functions as either Discrete or Continuous. Continuous Optimisation involves problems modelled using continuous variables. The requirement or classification of the variables as either discrete or continuous is in itself a criterion which influences the nature of the model. According to Hammer2000, the three notable branches of discrete optimization are: Combinatorial Optimization, which refers to problems on graphs, matroids and other discrete structures; Integer Programming which introduces additional integer variable criteria; Constraint programming in which the properties of a solution are stated for the computer in a declarative style rather than a sequence of steps.

Many real-world optimization problems involve discrete choices and can hence be modelled as discrete optimisation problem. Popular generalised examples are The Knapsack problem, The Traveling Salesman Problem, etc.

This thesis focuses on special kind of discrete optimisation problems which are represented as graphs. This thus transforms the problem into a graph problem in which several tools in graph theory can be applied. Zaválnij2020 presents the expressive power of using of k-Cliques as a modelling and problem-solving tool with results from Szabo2016, and some algorithms that can give significant improvement in speed for applications using this tool were also highlighted. The focus of this thesis is to implement and evaluate the performance of some of these algorithms, particularly those involving kernelization.

## 2.2 K-Clique problem

### 2.2.1 Complete Graph

An undirected graph is referred to as a Complete Graph if every arbitrary pair of distinct vertices is connected by a unique edge. For a directed graph, these vertex pairs are connected by pairs of unique edges. The number of vertices in such a graph is often denoted by the letter "k" Figure 1 depicts a list of k2, k3, up to k7 complete graphs.

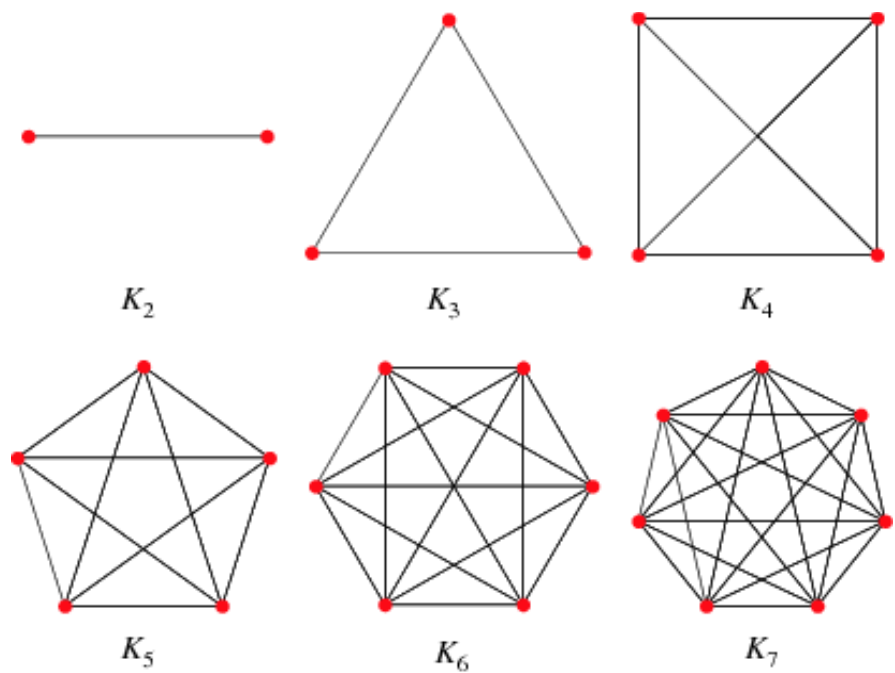

Figures 1: Complete Graphs. [6]

### 2.2.2 Induced Subgraph

When a new graph G' is formed from a subset of the vertices of another graph G and the edges connecting the subset of the vertices are retained, then the graph G' is called an induced subgraph of G. Such that G' is induced by V' and E' as in the figure 2.

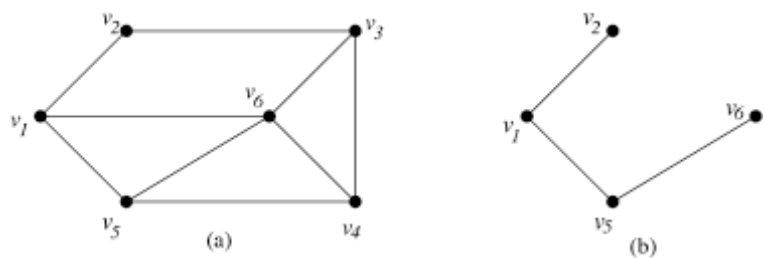

Figures 2: Induced Subgraph []

## 2.2.3 Clique

A clique of a graph G is an induced subgraph of G, which is a complete graph. So, applying the notation used for complete graphs, it is said that a k-Clique graph has k vertices. A k-Clique graph is also said to have a size of k. Figure 3 gives examples of cliques of k size.

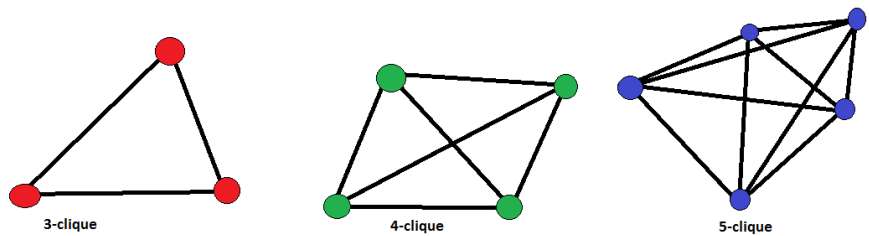

Figures 3: k-Cliques Graphs []

In computer science, the task of determining whether a particular graph contains a clique of a given size is referred to as the Clique or k-Clique problem. The task can be modelled in diverse ways mostly involving either finding the cliques or the information about the cliques. Some popular optimisation problems involving k-Cliques include listing all maximal k-Cliques, which cannot be extended; finding a maximum sized k-Clique in a graph; finding a maximum weighted k-Clique in a weighted graph.

Many problems in various fields can be modelled as a k-Clique problem such as Social Networks analysis [2], Bioinformatics [4], Electronics in IC design [3], Computational science and more.

The study of cliques has a long history in mathematics from Erdős & Szekeres (1935) and it continues to be an attractive are of research. For instance, the combinatorial problem which Paul Erdős would like to see solved, now labelled the Erdős–Faber–Lovász conjecture, remain unsolved.[10] In general, discrete optimisation problems are intrinsically difficult to solve and most of the techniques used to solved comes with additional complexity. As noted in Zaválnij2020, modelling these problems as graph introduces some practical challenges. The k-Clique problem has been proved to be one of the computationally hard problems [1] and was categorised among the Karp's 21 NP-complete problems in Karp1972.

**2.3     What makes a problem hard?**

This is one of questions in the computational complexity theory discipline where a question or problem a computer might be able to solve is categorised in terms of its resource usage. The resources considered may vary based on the context, but computational time seems to a common resource in most context. This is often referred to as the Computational Time Complexity, which approximates the compute time it takes for an algorithm to run. It basically counts the number of atomic operations performed within the algorithm. So, considering the naïve factorial algorithm below, it will perform the multiplication operation "n" times recursively. This algorithm is therefore said to have a time complexity denoted by O(N). This is the worst-case time complexity which the one often considered. In this case best case is the same as the worst case as there will be no multiplication executed which is when n = 0.

factorial (n):
|        if (n = 0) then return 1
|        otherwise return n * factorial(n-1)

In the case of the naïve factorial algorithm the computational time T(n) increase in proportion to the input size n, T(n) is thus bounded by n and the algorithm is said to have a Linear time complexity, O(N). For some algorithms it can be constant denoted as O(1). An algorithm has an exponential time complexity if T(n) is bounded by $O(2^{n^k})$ for some constant k.

A polynomial time complex algorithm is the one in which T(n) is bounded by a polynomial expression of its input size n. It is denoted by $O(n^k)$ for some positive constant k. The notion of polynomial time complexity has led into an interesting classification of problem as Complexity Class. This classification is often used to argue the efficiency of an algorithm. The most popular among these classes are: [11]

- P class: The complexity class of problems that can be solved on a deterministic Turing machine in polynomial time. In essence, these are problems for which a deterministic polynomial time algorithm exists.

- NP class: The complexity class of problems that can be solved on a non-deterministic Turing machine in polynomial time. The solution to these problems can be verified with proofs in polynomial time by a deterministic Turing machine.

A non-formal way to understand a Deterministic Turing machine is a computation in which given a particular input, a single output result is defined while a Non-Deterministic Turing machine is a computation in which given a particular input, more than one output results is defined. So, you cannot predetermine the output of a Non-Deterministic Turing machine.

The complexity class categorises a problem based on the speed of the fastest known algorithm for solving that problem, thus if a faster algorithm is discovered a problem may change class. By implication problems that takes more time to solved are considered harder than those that take lesser time. This can be generalised beyond the computation time to apply to order computer resources like memory.

Observing both definitions above it is obvious that if a problem can be solved deterministically in polynomial time, thus making it a P class problem. Its solution can be verified deterministically in polynomial time by simply solving the problem, this makes the problem also an NP class problem. So clearly P class problems are a subset of NP class problems.

While investigating the relationship between the members of a complexity class we seek for a function g(x) that can transform one member into another. If a member problem A can be transformed into another member problem B, then it is possible to deduce unknown properties of problem A from the properties of problem B and vice versa. This is known as Reducibility, such that if g(x) exists then A is reducible to B. [11] One important property preserved by reducibility is the complexity property.

The class of problems for which, given an element E of this class, any NP class problem can be reduced to it is known as NP-Hard Complexity class. These problems are at least as hard as any NP class problem but may lie outside the NP class set.

NP-Complete problems are those problems which are both NP-Hard and are also completely with the NP class.[12]

The question of whether the complexity class of P is a proper subset of the complexity class of NP (i.e., Is P = NP?) is still open. However, in a poll [13] most researcher in this field expects the answer to this question to be NO! This situation is often considered when presenting the complexity classifications because the outcome of this question has profound implications across various disciplines. Figure 4 depicts the relationships between the members of this classification considering both outcomes.

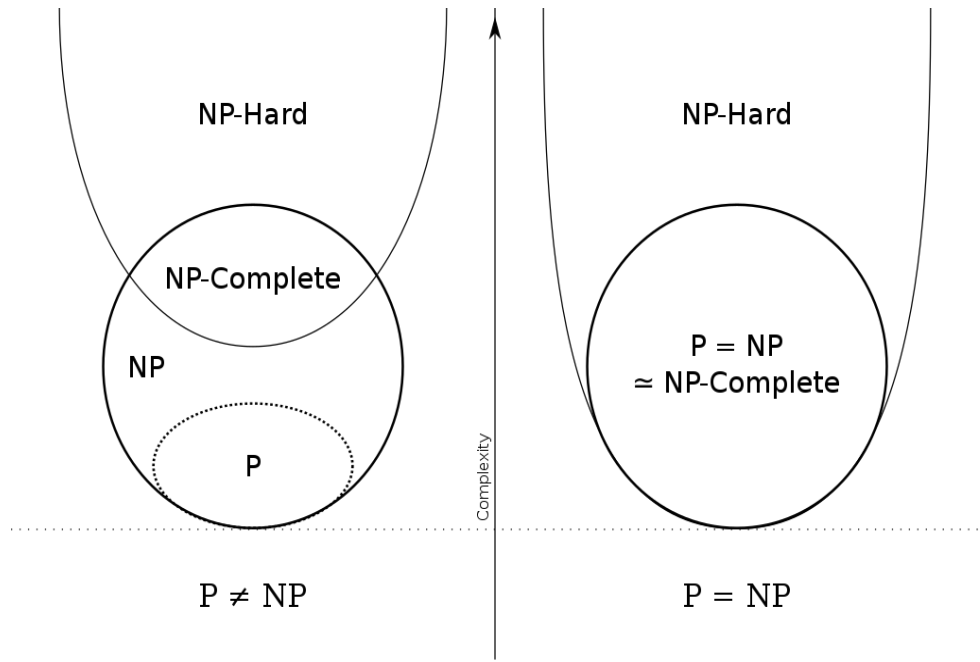

Figure 4: Euler diagram for P, NP, NP-complete, and NP-hard set of problems

## 2.4 Kernelization

To solve computationally hard problems, programmers use various kinds of heuristics in practice. Kernelization is a pre-processing technique used to reduce the instance size of computationally hard problems into its subset called Kernel.[14] It uses heuristics to find the Kernel such that the solution to the problem with the Kernel as input is also valid for the original input set or differ only by a linear transformation.

As Neeldhara et al noted, if we find the Kernel of an NP-hard problem and solved the problem in polynomial time, then would have proved that P = NP, which is expected to be unlikely.[14] The catch is that pre-processing algorithms like Kernelization are analysed within the context of parameterized complexity theory and not within the domain of classical complexity theory.

Parameterized complexity is a computer science field that extends the classical complexity theory by introducing a two-dimensional analysis of problems instances – one dimension used as usual for measuring the input-length, and the other used for measuring other structural-properties of the input.[15] This allows for a refined categorisation of tractable and intractable computational problems. In this context, a problem is said to be fixed-parameter tractable (FPT) if it has an algorithm running in time, $f(k)p(n)$ (FPT-time), where $f$ is any computable function solely in the parameter $k$, and $p(n)$ is a polynomial in the total input length $n$ [15]. Kernelization is an essential technique for designing FPT Algorithms. In an early published work H.L. Bodlaender et al states,

"A kernelization algorithm for a parameterized problem is a polynomial-time transformation that transforms any given instance to an equivalent instance of the same problem, with size and parameter bounded by a function of the parameter in the input. Typically, this is done using so-called reduction rules, which allow the safe reduction of the instance to an equivalent "smaller" instance. In this sense, kernelization can be viewed as polynomial-time pre-processing which has universal applicability, not only in the design of efficient FPT algorithms but also in the design of approximation and heuristic algorithms." [15]

# 3.  Parallel Architectures

This chapter gives a brief narration of the journey to modern processing units, focusing on events that necessitate the need for multiple processing units and the challenges that were faced and were created.

## 3.1  Multicore Processing

The Moore's law or observation is a phenomenon which states that after every two years the number of transistors cramped in a dense integrated circuit chip will double. It was first observed by Gordon Moore in 1965 and this prediction as shaped how CPUs have been designed. Encouraged by this prediction, CPU architects continuously cramp more transistors per area, essentially doubling the computing capability in clock speed every two years. As computers gets smaller and faster, more disciplines benefit by solving hard problems. The CPU architects would be further encouraged by another observation now known as the Denard scaling.

Dennard scaling is another observation proposed in 1974 by Robert H. Dennard et. al, it purports that the power density of transistors remains constant as transistors get smaller. So roughly, the power used by a transistor will also reduce as we reduce its size.[16] This is a clear encouragement to exploit the Moore's law and continue to increase the computation capability per die in terms of clock frequency. This was achieved to a great deal without significant increase in the overall circuit power consumption on the chip. However, as shown in figure 5, around 2005, the power seems to remain constant suggesting the end of Dennard Scaling. This means that CPU designers can no longer continue to increase the clock frequency on a single CPU. To circumvent this roadblock multiple CPU cores are then fabricated on a single chip. So, if Moore's law still holds, the more transistors available on the chip can be used to enable more CPU cores. We may be stalled at clock frequency per CPU core, but we will continue to have increase in the overall speed per chip. This is the birth of multicore processors.

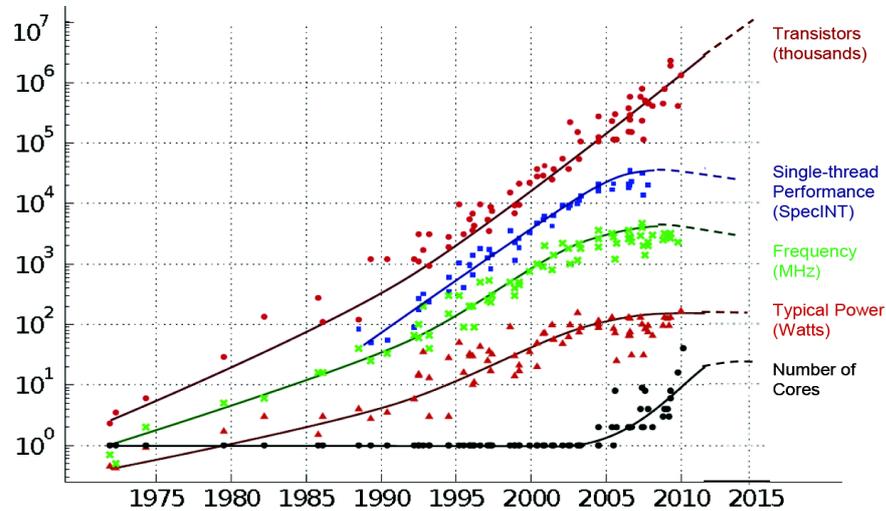

Figure 5: Graph showing of trends in various properties of CPUs over three decades [17]

The period of the single core computers is often referred to as the free lunch period [18] where algorithms are designed to exploit the simplicity of the Von Neumann architecture. In that era, to improve the speed of your program you must buy a faster hardware, if you already have the fastest hardware then you just wait two years for Moore's law to ripe and you get twice your speed by doing nothing but acquiring the fastest hardware. The free lunch era enabled many applications and even revived some dead disciplines like Artificial Intelligence because operations which are previously thought to be impossible then became possible due to increase in compute capability. However, with multicore systems the CPU is more complicated, and a faster CPU does not necessarily mean a faster program. You must design and implement your algorithm to take advantage of the multicore processor.

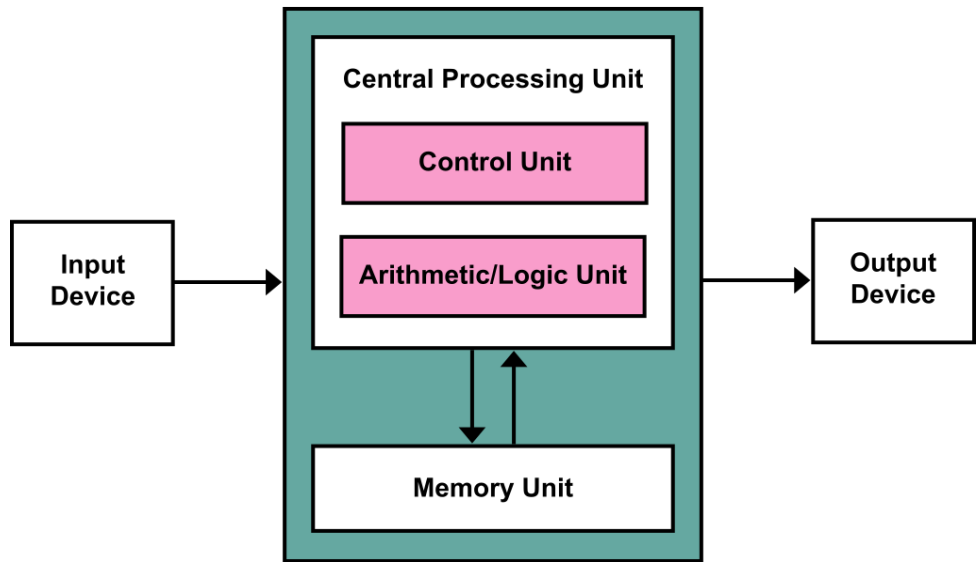

Figure 6: Von Neumann's Architecture [19]

### 3.2 Concurrency is not Parallelism.

Parallelism execution is doing many things at the same time such that each task is processed simultaneously by an independent processing unit. The goal here is to improve the throughput of the system.

Serial or sequential execution is doing many things one after the other by the same single processing unit. The key property of this model is that only one process entity is being executive by the system at any point in time while others are scheduled for execution serially.

Concurrent execution is doing many things such that they are all handled by one processing unit but access to the processing unit is scheduled not serially bought in a multiplexed fashion, such that these gives an impression of a simultaneous or concurrent execution. The goal of concurrency is to decrease the response time of the system, and this is what gives the impression of parallelism. The mechanism of this model is that there is a switch to a different task when the current task is not making useful progress. In a typical example, while waiting for an I/O task, the processing unit can switch to another useful task execution. The costs of switching contexts in concurrency often impairs the throughput of the system.

The concept of concurrency has been used even before the multicore era to enable multi-processing and multi-tasking in systems. In modern multicore architectures, it remains particularly useful in designing it practical parallelised system otherwise the system will only be capable of processing tasks that are all exclusively independent, these are often referred to as embarrassingly parallel tasks. Most real-world problems require some form of shared access to a system resource, mostly the memory, so an efficient concurrent scheduling is important. There lies the challenge of programming on a parallel architecture.

The current architecture of the CPU, though still largely best on the simple von Neumann architecture as depicted in figure 6, as evolved into a more complicated edifice. Figure 7 illustrates the memory hierarchy of a modern CPU architecture - the Intel core i7. The though the system bus is left out in this illustration, but the story is still the same. The modern CPU architecture is overly complicated. This matters for programmers because it further makes it more complicated to take full advantage of the computation capability of such complex hardware, the algorithm design and implementation need to be aware of these complexities.

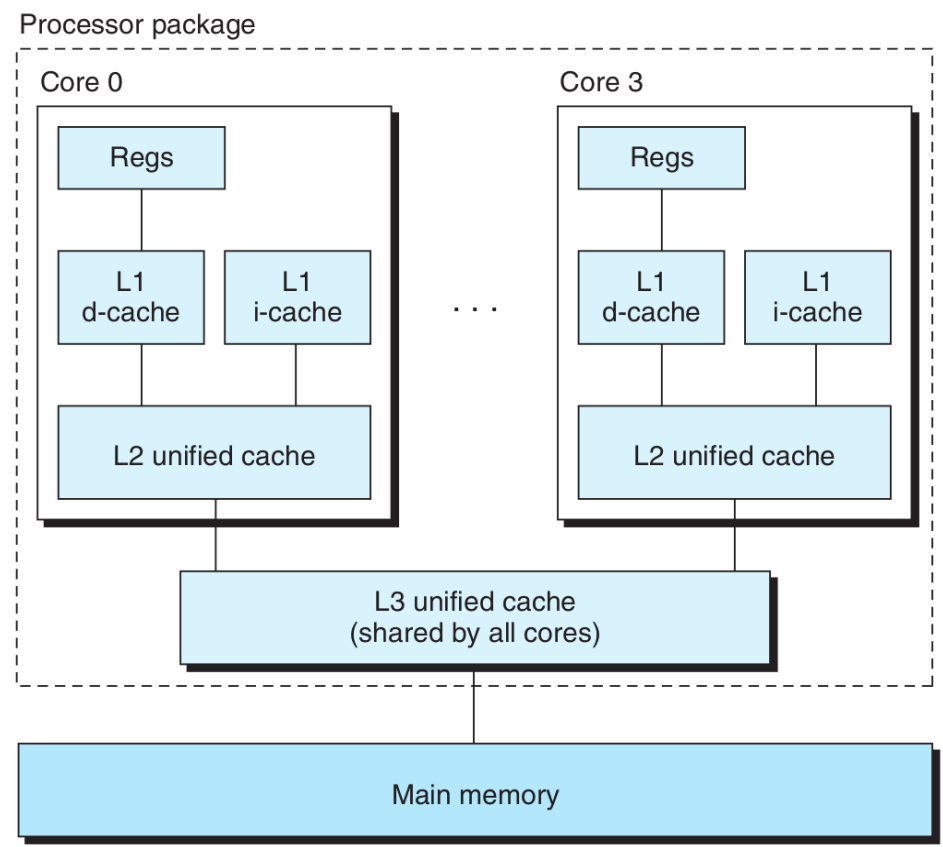

Figure 7: An abstract architecture of a multi-core processor emphasising the memory hierarchy (Intel core i7). [20]

## 3.3   Types of Parallelism

**Data parallelism:** the same operation or instruction is performed on a considerable chunk of data by different compute units at the same time. The original data is broken down into chunks which are then distributed to the available compute units.

**Task parallelism:** different compute units perform different operations or instructions all at the same time. Depending on the platform or organisation vocabulary this also be expressed as Thread-Level parallelism. In Thread-Level parallelism the thread is a unique compute unit for which its definition can differ across different companies.

**Pipeline parallelism:** instructions or operations assembled in serial such that the output of one instruction serves as input to the other, thus forming a linear pipeline. In this mode each individual compute unit is assigned a different pipeline. Data is usually streamed into the pipelines for maximum utilisation of the compute units by keeping them always busy.

**Instruction Level parallelism:** this type of parallelism is conceptually different from the ones previously discussed. This level of parallelism operates after assigning an operation or instruction to a processing unit. The goal is to execute the assigned operation in parallel on the hardware. This means that the hardware must support it and expose an interface to access this level of parallelism. An example are CPUs which support vector processing.

## 3.4   Flynn's taxonomy of parallel architecture

This is the most popular classification of computer architecture, and it was proposed by Michael J. Flynn more than five decades ago.[21]

**Single Instruction Single Data (SISD):** these is an architecture in which one instruction at a time is operating on each chunk or stream of data. This architecture involves no parallelism. It is typical of single core processor systems.

**Single Instruction Multiple Data (SIMD):** This architecture involves a single instruction operating on multiple different chunks or stream of data. This is common with vector processors, modern GPU (Graphics Processing Unit), and accelerators. The NVIDIA GPU architecture platform extends the idea of this model into SIMT (Single Instruction Multiple Thread).

**Multiple Instruction Single Data (MISD):** in this architecture, multiple instructions operate on one single data. This architecture is uncommon and will require all participating systems to agree on the result. This idea may be useful in fault tolerant systems e.g., space shuttle flight control computer.

**Multiple Instruction Multiple Data (MIMD):** Here multiple instructions operate on multiple data simultaneously. This may be implemented on one CPU or on different autonomous CPUs. An example of this architecture is a multi-core super scalar processor.

## 3.5  Amdahl's law

The computer scientist Gene Amdahl proposed what is now known as the Amdahl's law in 1967. It presents a formular for calculating the theoretical speedup achievable in latency of the execution of a task running on a processing unit at a fixed workload. It often used to predict how much speedup can be expected if multiple processing units are deployed on a task.

For example, if a task running on a single thread completes in 20 hours, but a portion of the task which cannot be parallelised consumes 1 hour of the total time, so we have 19 hours of execution that can be parallelised. The theoretical speedup can be calculated thus:

$$S(N)=1/((1-P) + (P/N))  [22]$$

- P is the proportion of a system or program that can be parallelised. (e.g., 19/20 = 0.95 in the above example)
- Note that (1-P) denotes the proportion that remains serial.
- N is the number of processing units.
- S(N) is the maximum theoretical speedup achievable on this task using N processing units.

It can be observed from the formular that, as the number of processing units N, grows the speedup tends towards 1/(1-P), which is 20 for the above example. It cannot go beyond this value; hence the speedup of any algorithm is limited by the total time spent on serial part of the algorithm. This implication of Amdahl's is incredibly significant because extracting parallelism is hard enough but even worse is that after the arduous work you are still limited by the serial section of your algorithm. This has result in the rise of specialised processing units tailored to a specific domain and can even sometimes be coupled with traditional CPUs in what has become known as heterogenous computing technique.[23]

## 4. Methodology

This chapter expounds on the background behind the algorithms implemented in this thesis. They are all based on previous and ongoing works like Szabo2019. The descriptions given are based on the approaches as highlighted in Zaválnij2020.

Consider the following two problems:
1. Given a finite simple graph G. What is the size of all maximum cliques in G.?
2. Given a finite simple graph G and given a positive integer k. Is there a k-Clique within the graph G?

The first problem is commonly referred to as the Maximum Clique problem which an NP-Hard problem, while the second problem defines the k-Clique problem which is an NP-complete problem.[7] The k-Clique problem is listed third in the Karp's original 21 NP-complete problems.[1] This thesis implements algorithms in Zaválnij2020, which focus on algorithms for exact solutions.

Based on their complexity classes, these two problems share some properties. Besides, a search for all maximum cliques will obviously include the search a clique of size k. So, an algorithm that solves one of these problems can also be used to solve the other. In particular, if we find an algorithm that solves the k-Clique problem, we can run this algorithm multiple times for different values of k in order to find all maximum cliques in the graph.[7]

As discussed earlier in Section 2, NP-complete problems are extremely hard to solve. Any algorithm for solving an NP-complete problem like the k-Clique problem, is bound to take exponential time in the worst case, unless P = NP! In practice, a direct application of the algorithm to the problem is not feasible. Instead, programmers employ several heuristic and brute force methods that are quite amenable to programming to tackle this class of problems. To further, reduce the problem space the technique of kernelization is growing in popularity. [7]

Kernelization is a pre-processing technique used to reduce the instance size of computationally hard problems into its subset called Kernel. It uses heuristics to find the Kernel such that the solution to the problem with the Kernel as input is also valid for the original input set or differ only by a linear transformation.

Zaválnij2020 categorised some of these kernelization algorithms into Structions, Colour indices and Dominance. This thesis focus on parallel implementation of some of these algorithms.

## 4.1  Structions

The algorithms in this category are based on Ebe1984 and Ale2003. Here a direct search for the cliques is done.

**Definition 4.1.1:**
Given a finite and simple graph G.
Let V = {1, …, n} is the set of vertices in G

For purpose of constructing the struction, node 1 in V is marked as the pivot node.
Let A = {a1, …, aI} be a set of the neighbours of node 1
Let B = {b1, …, bJ} be a set of non-neighbours of node 1

Such that: the set B is ordered so that b1 < … < bJ
We can then construct a new set C from the set B thus:

The set C contains the element c(bX, bY)
If:
- bX, bY is an element of B.
- The unordered edge {bX, bY} is an edge of G.
- X < Y

Hence the struction G' is define thus:
The set of nodes of the struction graph G' = the union of the set A and the set C such that:
- two nodes aI and aJ from the set A, are adjacent in G' only if the unordered pair {aI, aJ} is an edge of G.
- two nodes c(bX, bY) and c(bN, bM) from the set C, are adjacent in G' only if bX=bN and the ordered pair {bY, bM} is an edge of G.

**Lemma 1:** if CLQ' is a clique in G', then there is a clique CLQ in G such that |CLQ| = |CLQ'| + 1. A detail proof can be found in Zaválnij2020.

**Lemma 2:** if CLQ is a clique in G, then there is a clique CLQ' in G' such that |CLQ'| = |CLQ| - 1. A detail proof can be found in Zaválnij2020.

**Theorem 1:** as a corollary of Lemma 1 and 2, it can be said that a clique of the size of G' is one less than the size of a clique of the size of G.

### 4.2     Colour indices

Here instead of embarking on a direct search for the cliques, the original graph is searched for nodes or edges that can be deleted without affecting the cliques.

Concepts:

Graph labelling is the assignment of an identifier, usually integers to the edges or vertices of a graph.

Graph colouring is a special case of labelling in which the labels are referred to as colours (but does not have to be actual colours). The assignment of colours to graph elements (vertices/edges) is often subject to certain constraints.

Edge or Vertex colouring is the colouring done such that no two adjacent edge or vertices have the same colour. When this is achieved for a given graph, it is said that the colouring of that graph is Legal.

**Definition 4.2.1:**

Given a finite and simple graph G. It is legally coloured using k colours such that the each of the colours form a colour class. So, the colour classes are: $c_1, c_2, …, c_k$.

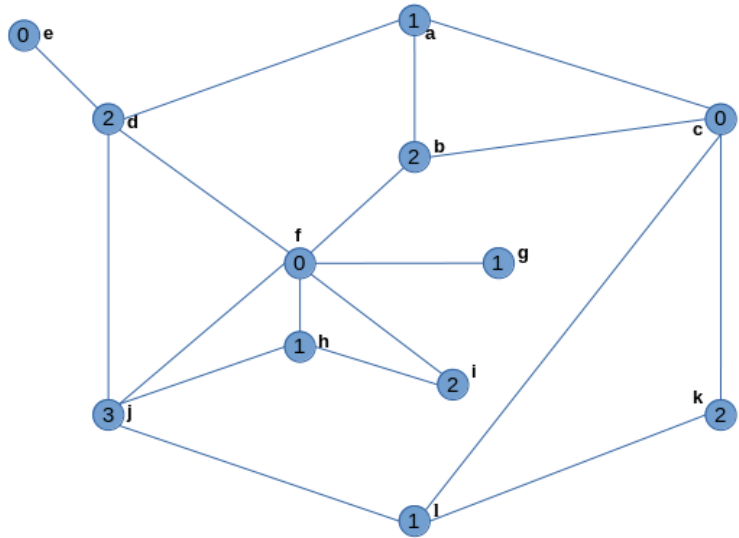

Figure 8: A graph of 12 vertices, leally coloured with 3 colours.

The Colour Index of a node v, in G, (with respect to a legal colouring of G) is the number of colour classes, that contains at least one node that is adjacent to v.

e.g.:

vertex a from figure 8 has colour index of 2: {d (2), b (2), c (0)} = {2, 0}

vertex d from figure 8 has colour index of 3: {a (1), e (0), j (3), f (0)} = {0, 1, 3}

**Remarks:**

If the Colour Index of a node v is less than k-1, then v cannot be a node of a K-clique in G. In essence, all nodes having a colour index of at most k-2 can be deleted without losing any k-clique.

The Colour Index of an edge e = {u, v} in G, (with respect to a legal colouring of G) is the number of colour classes, that contains at least one node that is adjacent to u and v simultaneously.

e.g.:

edge {a, d} from figure 8 has colour index of 0.

edge {h, i} from figure 8 has colour index of 1: {f (0)}

If the Colour Index of an edge e = {u, v} is less than k-2, then e = {u, v} cannot be an edge of a K-clique in G. So, e = {u, v} can be deleted when looking for a k-clique in G, however, we must not delete the nodes u and v.

## 4.3    Dominance

In this case too, instead of embarking on a direct search for the cliques, the original graph is searched for nodes or edges that can be deleted. Though in some cases the deletion affects the cliques, the effect is minimal. [7]

**Definitions 4.3.1:**
Let a and b be two distinct nodes of a finite and simple graph G. N(a) and N(b) are the set of the neighbours of a and b, respectively.
Then:
Node b is said to dominate node a
If a and b are not neighbours and $N(a) \subseteq N(b)$.

**Remarks:**
If node a is dominated by node b, then node a can be deleted from the graph during a search for a k-Clique. Although, there is a possibility of losing some k-Cliques during this reduction, not all k-Cliques will be lost.

**Definitions 4.3.2:**
Let a, u and b be three distinct nodes of a finite and simple graph G. N(a), N(u) and N(b) are the set of the neighbours of a, u, and b, respectively.

Then:
Edge {u, b} is said to dominate the edge {a, u}
If: b is NOT an element of $N(a) \cap N(b)$ and $N(a) \cap N(u) \subseteq N(u) \cap N(b)$

**Remarks:**
If the edge {a, u} is dominated by the edge {u, b}, then edge {a, u} can be deleted from the graph during a search for a k-Clique. However, the nodes a and u are not deleted.

**Definitions 4.3.3:**

Let x, y, u, and v be four distinct nodes of a finite and simple graph G. N(a), N(u) and N(b) are the set of the neighbours of a, u, and b, respectively.
 If the pairs {x, y} and {u, v} are edges of G,

Then:
Edge {u, v} is said to dominate the edge {x, y}
If:
- either {u, x} or {u, y} is not an edge of G,
- and either {v, x} or {v, y} is not an edge of G,
- and $N(x) \cap N(y) \subseteq N(u) \cap N(v)$

**Remarks:**

If the edge {x, y} is dominated by the edge {u, v}, then edge {x, y} can be deleted from the graph during a search for a k-Clique. However, the nodes x and y are not deleted.

## 4.4   Implementation

The implementation of these algorithms is done using C++. Effort is made to stick to modern version of C++ especially C++17 which has a better support for writing parallel code than previous versions. Also, the use of smart pointers to manage lifecycle of pointers is also desirable. However, this wish would fall apart because the version of CUDA available for testing this thesis does not support C++ natively. Effort was made to downgrade the code to at least C++11 the first version of what is today tagged modern C++.

The input space for the algorithm is a graph. The goal of kernelization is to reduce this input space significantly and to achieve this efficiently. Besides the obvious reasons an algorithm that converts fast is desirable because it is not in all cases that a significant reduction of the input space is achievable, if at all. It desirable to get this information on time begin actual process of solving the problem.

4.4.1 Graph Representation

There are several ways to represent the graph and the choice is often determined by the nature of the algorithms. Table 1 shows a comparison between Adjacent Matrix Edge List and Adjacent List on three operations performed frequently in all the implemented algorithms. The operation IsEdge() occurs far more than any of the other operations, so with an O (1) complexity, the choice of Adjacent Matrix seems obvious. Adjacent Matrix offers even more advantages because operations like finding intersections of neighbours can be reduced to a simple vector addition. The same thing applies to checking is a set of vertices are a subset of other vertices. Reducing these operations into vector operation helps take advantage of underlying parallelism and this is where the major speedups were achieved.

|                 | **IsEdge()** | **ListAllEdges()** | **ListNeighbours(v)** |
|-----------------|--------------|--------------------|------------------------|
| **Adjacent Matrix** | O (1)        | $O(|V|^2)$         | $O(|V|)$               |
| **Edge List**   | $O(|E|)$     | $O(|E|)$           | $O(|E|)$               |
| **Adjacent List** | O(degree)  | $O(|E|)$           | O(degree)              |

Table1: Comparison of common operations on 3 different graph representations.

4.4.2 Random Graph generation

All the algorithms expect an input graph that is at least undirected, simple, finite and connected. To ease the testing a random graph is generated which may or may not be saved. However, the final analysis is done using a few save graphs randomly generated. The random graph function takes as input the sizes of both vertices and edges. The longest path through graph is first constructed by a sequential visit to all the vertices. The remaining edges are then generated in a pseudo-random fashion.

4.4.3 Dominance Algorithms

Multiple parallel versions of each algorithm were implemented in addition to the serial version. OpenMP and NVIDIA CUDA parallel frameworks were used for the parallel implementations. Though more than three algorithms were implemented, the section use the three dominance algorithms highlighted in Zaválnij2020 to illustrate the chosen interpretations and the implementation decisions.

Algorithm 1 is based on definition 4.3.1 and this is the only algorithm where the dominated vertices are deleted. It often results in deleting some but not all the k-Cliques in the graph. Algorithm 2 is based on definition 4.3.2 while Algorithm 3 is based on definition 4.3.3. There is a similarity in the preconditions of both Algorithm 2 and Algorithm 3 with requirement of a distinct set of vertices.

**Algorithm 1** Dominance Algorithm 1
1: Initialize $G(v)$, for all $v \in V$ (*The set of vertices of Graph, G*)
2: **for all** *node* $\in V$ **do**
3:    $N(node) \leftarrow$ *All neighbours of node*
4:    **for all** $a \in N(node)$ **do**
5:       $N(v) \leftarrow$ *All neighbours of a*
6:       **if** $N(a) \cap N(node)$ **then**
7:          delete a.
8:       **end if**
9:    **end for**
10: **end for**

A choice was made use to fulfil this requirement using different approaches in both algorithms. One approach uses a permutation algorithm to generate a set of all permutation of the number of required vertices, three for Algorithm 2 and four for Algorithm 3. It is possible to incorporate the algorithm within the search for these permutations and one implementation did this exactly, it uses a non-recursive permutation algorithm.[26] However, another approach that may be more parallelisable is to calculate the number of permutations required, then we can use algorithms that gives permutations in lexicographic order like the one proposed by Edger Dijkstra in Dijkstra76 to generate a specific permutation set at a particular position.

This means if as much thread as the number of permutations is spawned, since threads are numbered serially, it possible for each thread to generate its own permutation set and execute the algorithm. The algorithm itself has a linear time complexity if the intersection and subset operations are vectorised, the bulk of the time is used searching for the right permutation. Parallelising this permutation search is likely to yield a huge improvement.

**Algorithm 2** Dominance Algorithm 2
1: Initialize $G(v)$, for all $v \in V$ (*The set of vertices of Graph, G*)

```
 2: for all a Є V do
 3:   N(a) ← All neighbours of a
 4:   for all u Є N(a) do
 5:     N(u) ← All neighbours of u
 6:    for all b Є N(u) do
 8:      if (b == a) OR {u, b} not Є E (The set of edges of Graph, G) then
 9:        continue; //pick another vertex b.
10:      end if
11:
12:      N(b) ← All neighbours of b
13:      condition1 ← b Є (N(a) ∩ N(u))
14:      condition2 ← (N(a)∩ N(u)) ⊆ (N(a)∩ N(u))
15:
16:      if condition1 && condition2 then
17:        delete {a, b}; //only the edge is deleted not the nodes.
18:        break; //pick another vertex u.
19:      end if
20:    end for
21:   end for
22: end for
```

**Algorithm 3** Dominance Algorithm 3

```
 1: Initialize G(v), for all v Є V (The set of vertices of Graph, G)
 2: Initialize PermuatationSet(V) ← GetPermuatationSet(4, V)
 3: for all {x, y, u, v} in each set Є PermuatationSet(V) do
 4:    condition1 ← {x, y} && {u, v} Є E (The set of edges of Graph, G)
 5:    condition2 ← {u, x} || {u, y} not Є E
 6:    condition3 ← {v, x} || {v, y} not Є E
 7:    condition4 ← (N(x)∩ N(y)) ⊆ (N(u)∩ N(v))
 8:
 9:    if condition1 && condition2 && condition3 && condition4 then
10:       delete {x, y}; //only the edge is deleted not the nodes.
11:    end if
12: end for
```

# 5. Analysis and Evaluation

The implementations were executed on an Intel Core i5 processor. Appendix 1 shows the result of the evaluation of six out of the implemented algorithm. The application generates and saves a series of ten random graphs, each of which is then run though each of the algorithms. Initial analysis shows some of the algorithm would not converge within reasonable time on a typical laptop processor even for a graph with number of vertices as low as twenty. Algorithm 3 for instance run for more than a day and still did not converge for a vertex size of twenty.

The evaluated graphs have vertex size range between 100 and 1900. An arbitrary formular is used to compute the number of edges as a function of the number of vertices. It is the assignment of edges which are randomised; however, they are not fully randomised so that some vertices have more tendency of having multiple edges while still avoiding loops and parallel edges.

The table in Appendix 1 shows "Process Type" which identifies the algorithm and which implementation it is.

SERIAL EdgeDom v1 – serial implementation of a version of dominance algorithm.

OpenMP EdgeDom v1 – OpenMP implementation of SERIAL EdgeDom v1

SERIAL DOM. ALG-1 – serial implementation of Algorithm 1 above.

OpenMP DOM. ALG-1 – OpenMP implementation of Algorithm 1 above.

SERIAL DOM. ALG-2 – serial implementation of Algorithm 1 above.

OpenMP DOM. ALG-2 – serial implementation of Algorithm 1 above.

Other fields are "Num. of Vertices", "Num. of Edges", "Num. of Deleted Edges", "Num. of Deleted Vertices" and "Duration". These values are calculated for each implementation execution on each of the ten graphs. The graphs have vertex size: 100, 300, 500, 700, …, 1500, 1700, 1900.

## 5.1 Quantitative Analysis

The number of vertices and the number of edges as a correlation as stated earlier, so it no surprise that the graph as shown in figure 9, of their relation to the duration of execution for each algorithm looks the same. However, there are two outliers, these are related the two implementations of "EdgeDom" algorithm.

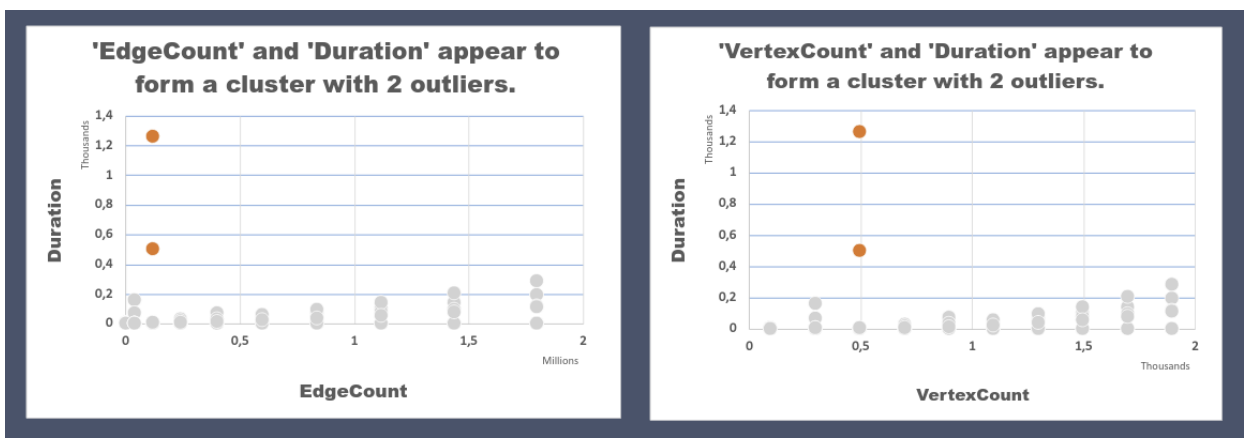

Figure 9: Scatter plot of Duration versus number of Vertices and number of Edges

"DOM. ALG-2" is the only algorithm is to be blamed for the outliers in the scatter plot analysis in figure 10. The plot visualises the rate at which the algorithms were able delete the edges and vertices. Here again there is a correlation between the plots, and it can be said that the plot for Duration versus Deleted Edges is contained in the other plot which is about deleted edges. The reason for this is that "DOM. ALG-1" as elaborated in Definitions 4.3.1 will delete dominated vertices along with the edges connected to it. This accounts for the fastest deletes in both edges and vertices.

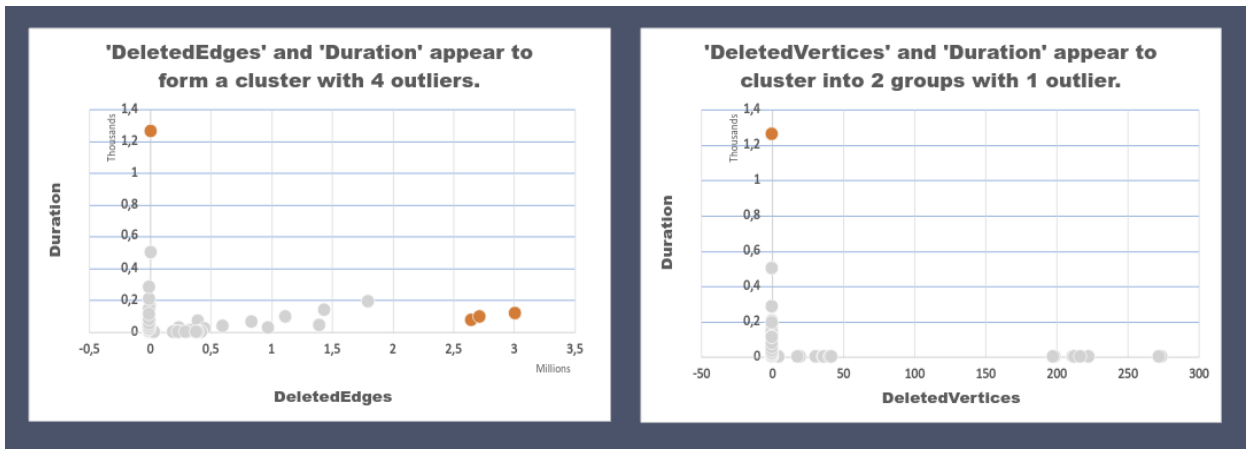

Figure 10: Scatter plot of Duration versus number of deleted Vertices and Edges

Figure 11 and 12 gives a deeper illustration of the details behind the earlier given analysis. Perhaps, the most significant part is that for all the algorithms the parallelised verion achieve sometimes more than 50% improvement over the serial version. As previously explained in section 3 of this thesis, parallelism is not for free. In this case, I am convinced that the parallelised versions have benefited from the reduction of some for-loop operations into a single vector operation which is then handled with Instruction-Level parallelism at least in the case of Clang++ which is the compiler used during the evaluation. Clang/LLVM has its own OpenMP implementation with exhibits this behaviour.

The EdgeDom algorithm as shown in figure 11 takes more time than any other algorithm and this is true for the OpenMP implementation. However, as depicted in figure 12, it is capable of producing the smallest kernel since it deleted the highest number of edges. Its OpenMP version is not only significantly faster, but it also produces a better kernel. The goal of kernel is to reduce the input graph significant within a short duration.

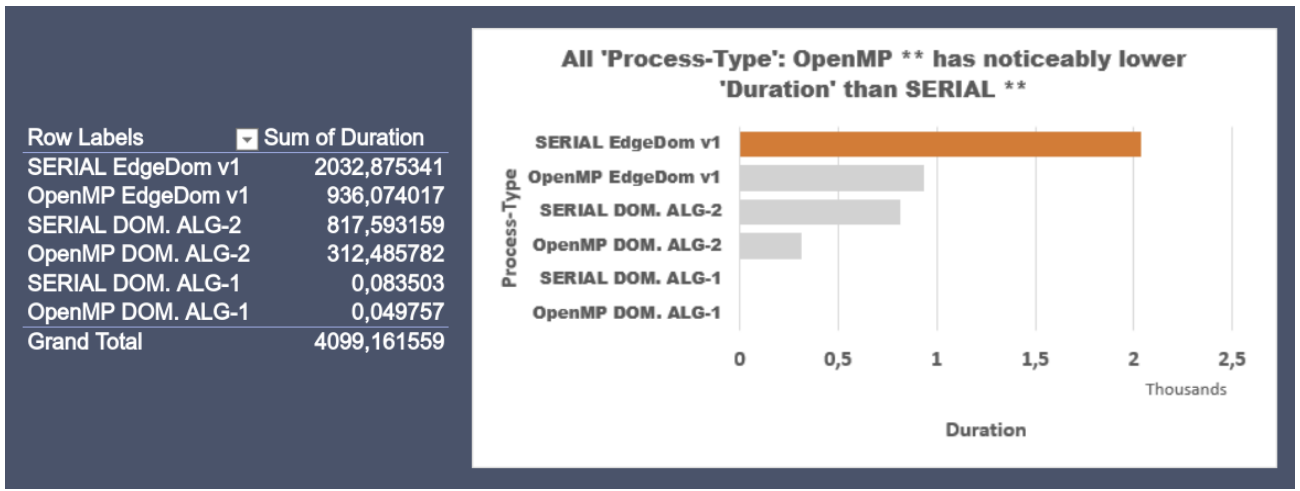

Figure 11: Data and Graphical analysis of the duration per implementation

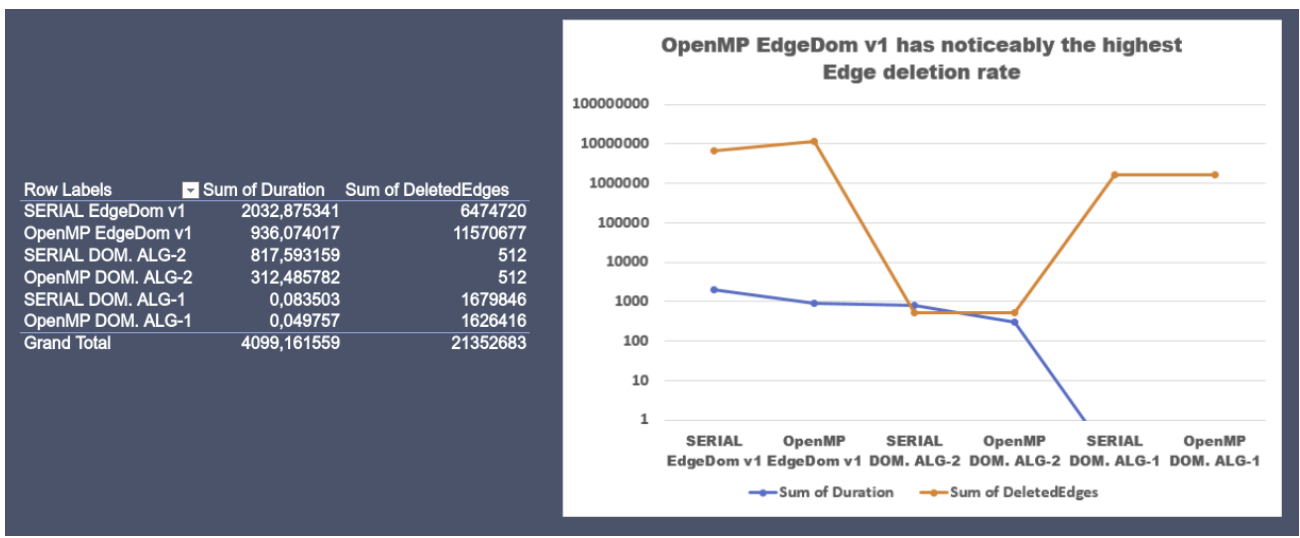

Figure 12: Data and Graphical analysis of the duration vs Edge Deletion per implementation

## 5.2     CUDA Implementation

The OpenMP implementation has clearly shown that it possible to harness parallelism on the CPU to aggressively extend the range of feasibility of kernelization algorithm. This motivates the evaluation of GPU or other accelerators which particularly designed for massive parallelism. Though the GPU diverse a great deal from the CPU, especially in terms of their memory model, the program design process is conceptually the same.

This evaluation does not contain the CUDA implementation of these algorithms, however, the initial implementation showed promising results. Unfortunately, due to logistic reasons, initial access to a suitable GPU was cut short, however the implementation at its current state is included in the source code accompanying this thesis.

## 5.3     Reproducing the evaluation

To reproduce the evaluation data used in this analysis, the accompanying code will need to be compiled. The code has been tested with both Clang++ and GCC compiler. It is advisable to use the latest versions or at least a version not older than five years. This is because both compilers now come with the OpenMP library bundled. The application also needs CMake but any build system of choice can be used. For convenience, I have generated the builds for both Eclipse and CodeBlocks IDEs they are in the directories "/eclipse-build" and "/codeblocks-build" respectively. It should be a matter of opening the project inside that folder within the respective IDE.

The graph used for this evaluation are generated randomly and save in the folder "/data/input". The executable file is "dominate", if this executed without any parameter the exact evaluation will be performed using the saved data. This will print out the result like appendix 1 on to the console. It will also save the same result as a csv file into the directory "/data/saved results" as a file called "result_.csv". Please, note that this file is overwritten at every execution.

There are two other options for executing the application:

- To generate a single graph and test with all algorithms:

$ dominate <number_of_vertices> <number_of_edges>

- To generate load a single graph from a file and test with all algorithms:

$ dominate <number_of_vertices> <path_to_the _graph_file>

Please, note that the application expects the graph to be saved as an adjacent matrix with each value separated by a newline.

## 6. Conclusion and Further Research

This thesis has focused on the implementation of a few kernelization algorithms on parallel architectures with the goal of harnessing parallelism on these platforms to aggressively extend the range of feasibility of these kernelization methods. My parallel implementations achieve over 50% improvement in efficiency for at least one of these algorithms. This is attained not just in terms of speed but it is also able to produce a smaller kernel. With growing interest in kernelization method, the results achieved in this thesis should serve as motivation to seek for more ways to improve the performance of these algorithm. The hope is that this will encourage further work on not just a proper GPU implementation but also on other accelerators or FPGAs.

# 8. List of Figures

# 9. List of Tables

# 10. Appendix

## 10.1 Appendix 1

Results Data:

| Process-Type | VertexCount | EdgeCount | DeletedEdges | DeletedVertices | Duration |
|---|---|---|---|---|---|
| SERIAL EdgeDom v1 | 100 | 4750 | 882 | 0 | 2,268885 |
| OpenMP EdgeDom v1 | 100 | 4750 | 550 | 0 | 0,944391 |
| SERIAL DOM. ALG-1 | 100 | 4750 | 457 | 5 | 0,000092 |
| OpenMP DOM. ALG-1 | 100 | 4750 | 454 | 5 | 0,000036 |
| SERIAL DOM. ALG-2 | 100 | 4750 | 8 | 0 | 0,042017 |
| OpenMP DOM. ALG-2 | 100 | 4750 | 8 | 0 | 0,016114 |
| SERIAL EdgeDom v1 | 300 | 44250 | 6791 | 0 | 156,374205 |
| OpenMP EdgeDom v1 | 300 | 44250 | 6414 | 0 | 65,700119 |
| SERIAL DOM. ALG-1 | 300 | 44250 | 5679 | 20 | 0,000703 |
| OpenMP DOM. ALG-1 | 300 | 44250 | 5131 | 18 | 0,000248 |
| SERIAL DOM. ALG-2 | 300 | 44250 | 9 | 0 | 1,120004 |
| OpenMP DOM. ALG-2 | 300 | 44250 | 9 | 0 | 0,43267 |
| SERIAL EdgeDom v1 | 500 | 123750 | 14702 | 0 | 1257,550175 |
| OpenMP EdgeDom v1 | 500 | 123750 | 15092 | 0 | 499,983185 |
| SERIAL DOM. ALG-1 | 500 | 123750 | 15778 | 33 | 0,001915 |
| OpenMP DOM. ALG-1 | 500 | 123750 | 14856 | 31 | 0,000677 |
| SERIAL DOM. ALG-2 | 500 | 123750 | 492 | 0 | 5,156118 |
| OpenMP DOM. ALG-2 | 500 | 123750 | 492 | 0 | 1,981604 |
| SERIAL EdgeDom v1 | 700 | 243250 | 243249 | 0 | 26,81928 |
| OpenMP EdgeDom v1 | 700 | 243250 | 332601 | 0 | 7,323553 |
| SERIAL DOM. ALG-1 | 700 | 243250 | 24384 | 36 | 0,003077 |
| OpenMP DOM. ALG-1 | 700 | 243250 | 25042 | 37 | 0,001268 |
| SERIAL DOM. ALG-2 | 700 | 243250 | 0 | 0 | 13,986554 |
| OpenMP DOM. ALG-2 | 700 | 243250 | 0 | 0 | 5,300587 |
| SERIAL EdgeDom v1 | 900 | 402750 | 402748 | 0 | 69,206435 |
| OpenMP EdgeDom v1 | 900 | 402750 | 452669 | 0 | 16,674065 |
| SERIAL DOM. ALG-1 | 900 | 402750 | 35895 | 41 | 0,004821 |
| OpenMP DOM. ALG-1 | 900 | 402750 | 36748 | 42 | 0,003149 |
| SERIAL DOM. ALG-2 | 900 | 402750 | 0 | 0 | 29,82921 |
| OpenMP DOM. ALG-2 | 900 | 402750 | 0 | 0 | 11,302872 |
| SERIAL EdgeDom v1 | 1100 | 602250 | 602352 | 0 | 36,980641 |
| OpenMP EdgeDom v1 | 1100 | 602250 | 979395 | 0 | 22,705414 |
| SERIAL DOM. ALG-1 | 1100 | 602250 | 210911 | 213 | 0,007788 |
| OpenMP DOM. ALG-1 | 1100 | 602250 | 198438 | 199 | 0,004652 |
| SERIAL DOM. ALG-2 | 1100 | 602250 | 0 | 0 | 54,293576 |
| OpenMP DOM. ALG-2 | 1100 | 602250 | 0 | 0 | 20,520641 |
| SERIAL EdgeDom v1 | 1300 | 841750 | 841750 | 0 | 60,543315 |
| OpenMP EdgeDom v1 | 1300 | 841750 | 1398317 | 0 | 41,709545 |
| SERIAL DOM. ALG-1 | 1300 | 841750 | 252430 | 212 | 0,010721 |
| OpenMP DOM. ALG-1 | 1300 | 841750 | 237141 | 198 | 0,006701 |
| SERIAL DOM. ALG-2 | 1300 | 841750 | 3 | 0 | 89,899263 |
| OpenMP DOM. ALG-2 | 1300 | 841750 | 3 | 0 | 34,170231 |
| SERIAL EdgeDom v1 | 1500 | 1121250 | 1121249 | 0 | 95,596979 |
| OpenMP EdgeDom v1 | 1500 | 1121250 | 2649519 | 0 | 74,188202 |
| SERIAL DOM. ALG-1 | 1500 | 1121250 | 308859 | 223 | 0,014442 |
| OpenMP DOM. ALG-1 | 1500 | 1121250 | 296066 | 213 | 0,008504 |
| SERIAL DOM. ALG-2 | 1500 | 1121250 | 0 | 0 | 138,703639 |
| OpenMP DOM. ALG-2 | 1500 | 1121250 | 0 | 0 | 52,303856 |
| SERIAL EdgeDom v1 | 1700 | 1440750 | 1440748 | 0 | 137,611634 |
| OpenMP EdgeDom v1 | 1700 | 1440750 | 2720125 | 0 | 91,063942 |
| SERIAL DOM. ALG-1 | 1700 | 1440750 | 427335 | 274 | 0,018145 |
| OpenMP DOM. ALG-1 | 1700 | 1440750 | 424486 | 272 | 0,012602 |
| SERIAL DOM. ALG-2 | 1700 | 1440750 | 0 | 0 | 202,870046 |

| Algorithm | | | | | |
|---|---|---|---|---|---|
| OpenMP DOM. ALG-2 | 1700 | 1440750 | 0 | 0 | 76,184761 |
| SERIAL EdgeDom v1 | 1900 | 1800250 | 1800249 | 0 | 189,923792 |
| OpenMP EdgeDom v1 | 1900 | 1800250 | 3015995 | 0 | 115,781601 |
| SERIAL DOM. ALG-1 | 1900 | 1800250 | 398118 | 223 | 0,021799 |
| OpenMP DOM. ALG-1 | 1900 | 1800250 | 388054 | 217 | 0,01192 |
| SERIAL DOM. ALG-2 | 1900 | 1800250 | 0 | 0 | 281,692732 |
| OpenMP DOM. ALG-2 | 1900 | 1800250 | 0 | 0 | 110,272446 |

## 10.2  Appendix2

Link to source code:

**https://gitlab.com/saheed.mik/kCliqueKernelize**